\def\year{2022}\relax
\documentclass[letterpaper]{article} % DO NOT CHANGE THIS
\usepackage{aaai22}  % DO NOT CHANGE THIS
\usepackage{times}  % DO NOT CHANGE THIS
\usepackage{helvet}  % DO NOT CHANGE THIS
\usepackage{courier}  % DO NOT CHANGE THIS
\usepackage[hyphens]{url}  % DO NOT CHANGE THIS
\usepackage{graphicx} % DO NOT CHANGE THIS
\usepackage{natbib}  % DO NOT CHANGE THIS AND DO NOT ADD ANY OPTIONS TO IT
\usepackage{caption} % DO NOT CHANGE THIS AND DO NOT ADD ANY OPTIONS TO IT
\DeclareCaptionStyle{ruled}{labelfont=normalfont,labelsep=colon,strut=off} % DO NOT CHANGE THIS
\usepackage{hyphenat}
\usepackage{algorithm}
\usepackage{algorithmic}

\usepackage{newfloat}
\usepackage{listings}
\floatstyle{ruled}
\newfloat{listing}{tb}{lst}{}
\floatname{listing}{Listing}
\title{Discovering Transition Pathways Towards Coviability \\
	with Machine Learning }
\author{
    Laure Berti-Equille\textsuperscript{\rm 1},
    Rafael L. G. Raimundo\textsuperscript{\rm 2}
}
\affiliations{
    \textsuperscript{\rm 1}IRD, ESPACE-DEV\\

    500, rue Jean-Fran\c cois Breton\\
    34093 Montpellier cedex 03, FRance\\
    laure.berti@ird.fr \\
     \textsuperscript{\rm 2}
     Federal University of Para\'iba - Campus IV \\
     Dept. of Engineering and Environment, Centre for Applied Science and Education \\
      Av. Santa Isabel s/n, Rio Tinto, PB, Brasil\\
     rafael.raimundo@academico.ufpb.br
     
}

\begin{document}

\maketitle

\begin{abstract}
	Coviability refers to the multiple socio-ecological arrangements and governance structures under which humans and nature can coexist in functional, fair, and persistent ways. Transitioning to a coviable state in environmentally degraded and socially vulnerable territories  is challenging. This paper presents an ongoing French-Brazilian joint research project combining machine learning, agroecology, and social sciences to discover coviability pathways that can be adopted and implemented by local populations in the North-East region of Brazil.
\end{abstract}

\section{Context}
The long-standing debate on how human societies can ensure the equitable use of common natural resources, conserve biodiversity and ecosystem functioning, and promote individual freedom and collective action \cite{Ostrom2011} has been developed into several schools of thought that define sustainability in different ways depending on their relative emphasis on ecological, socio-economic and political processes (see Fisher et al. (2007) for a comprehensive review). One of such schools is founded on the emerging concept-paradigm of {\bf socio-ecological coviability}, which refers to the multiple ways by which we can achieve a functional and persistent relationship between humans and non-humans under specific regulations and constraints (Barrière et al., 2019).

This paper presents our ongoing French-Brazilian collaborative project initiated in 2021 which aims at: (1) Establishing a diagnosis of socio-ecological coviability for several sites of interest in {\it Nordeste}, the North-East region of Brazil (in the states of Para\'iba, Cear\'a, Pernambuco, and Rio Grande do Norte respectively known for their biodiversity hotspots, droughts and vulnerabilities to climate change) using advanced data science techniques for multisource and multimodal data fusion and (2) Finding transition pathways towards coviability equilibrium using machine learning techniques. Data collected in situ by scientists, ecologists, and local actors combined with volunteered information, pictures from smart-phones, and data available on-line from satellite imagery, social media, surveys, etc. can be used to compute various coviability indicators of interest for the local actors. These indicators are useful to characterize and monitor the socio-ecological coviability status along various dimensions of anthropization, human welfare, ecological and biodiversity balance, and ecosystem intactness and vulnerabilities.

Among an infinite space of possible transition strategies, AI technologies and reinforcement learning (RL) in particular is applied to propose estimations and predictions of potential coviability pathways (or trajectories) for transitioning from a diagnosis state to an ideal coviability target (or intended view) defined by the user of the technology. Using computational models, our project aims to predict alternative solutions for coviability transitions at different time horizons and various spatial and temporal granularity levels. It will also reflect on the past and expose what would have happened had reality been different. The purpose of our research results is to be used by the local actors to be better informed and to improve the governance of territories facing environmental degradation and ecological vulnerability. 

Our consortium is composed of ten institutions (57 permanent researchers) in France and Brazil, working in interdisciplinary collaborations in environmental sciences (agroecology, hydrology), informatics (data science, remote sensing, geomatics, and machine learning) and social sciences (anthropology, social science, economics) lead by the Federal University of Para\`iba in Brazil and by IRD ESPACE-DEV research center in France. It involves four Brazilian universities (Federal University of Para\`iba (UFPB), Federal University of Rio Grande do Norte (UFRN), University of Sao Paulo (USP), Federal University of Pernambuco), four IRD\footnote{\tiny\url{https://en.ird.fr/}} research units (UMR ESPACE DEV, IMBE, PALOC, UMMISCO) in France, FUNCEME (Funda\c cao Cearense de Meteorologia e Recursos H\'idricos), GBIF France and Brazil, and three collaborators: INCT OndaCBC\footnote{\tiny\url{https://www.ondacbc.com.br/?lang=en}}   and INCT Odisseia\footnote{\tiny\url{http://inct-odisseia.i3gs.org/}} , and CEMAVE/ICMBio\footnote{\tiny\url{https://www.icmbio.gov.br/cemave/}}. Our joint research outcomes are intended to have deep societal, scientific and economical impacts, and be used to guide global conservation and restoration strategies toward the Nature positive 2050 target (Locke et al., 2020).

\begin{figure*}[t]
	\centering
	\includegraphics[height=6.7cm,width=14cm]{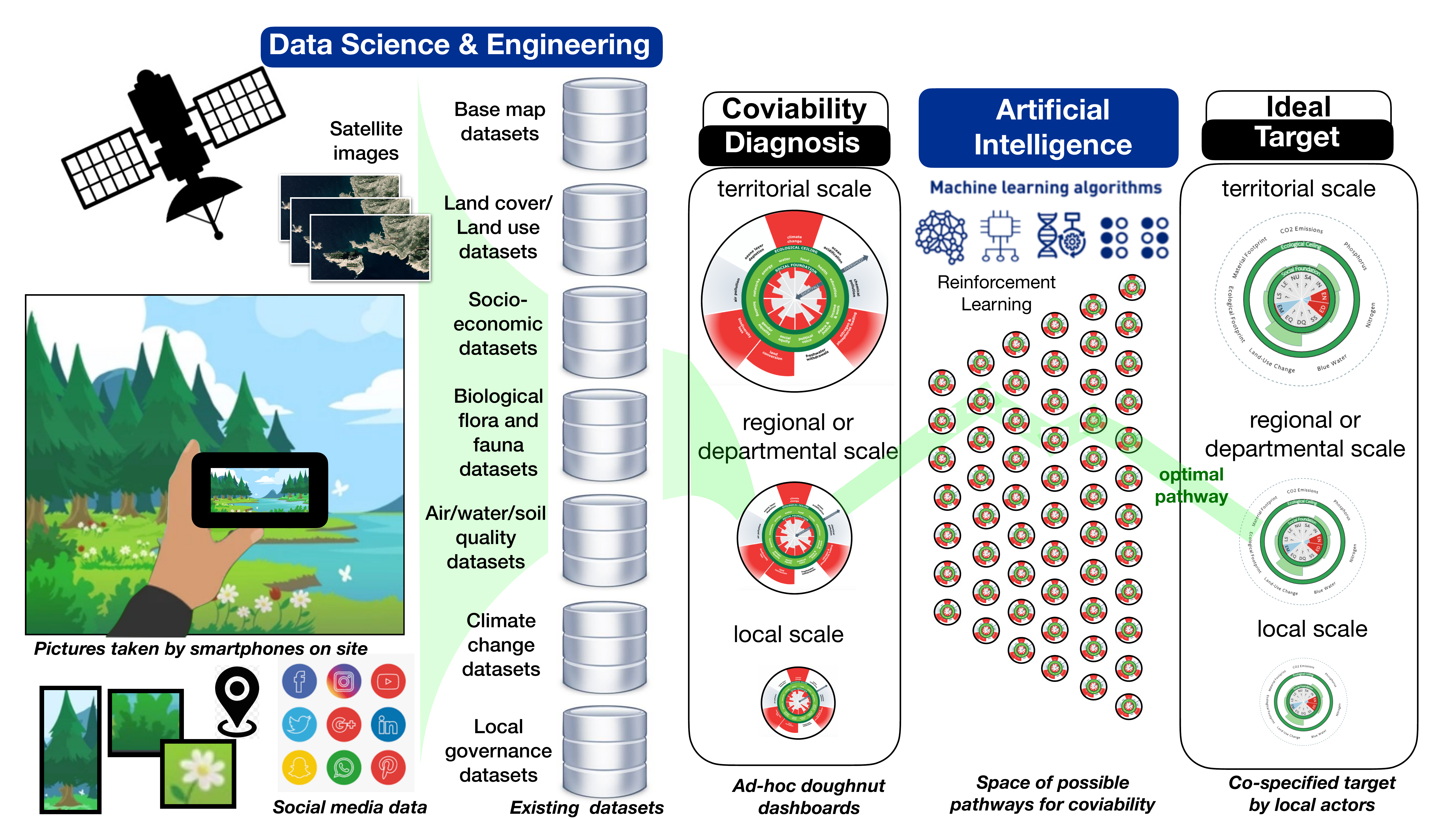}
	\caption{Overview of our Approach}
	\label{fig1}
\end{figure*}

\section{Overview of our Approach} 

The lack of coviability indicators on Brazilian territories often prevents local managers, farmers, and administrators from correctly establishing a diagnosis, monitoring its evolution, planning daily and seasonal  activities of the natural areas with considering carefully information about the life cycle of flora and fauna species living in the area (reproduction period, rarity, disturbance tolerance, bird migration key zones, etc.) and the intactness of their habitats to better preserve and restore the milieu. Without such indicators at various spatial and temporal scales, adapting the governance and justifying some regulations of the territory is a difficult task for the local actors. Relevant sources of information are usually not or under-exploited; eco-tourists, farmers, and populations are not informed of their impact on nature and they may not understand and accept such regulations. Global abrupt disconnection from nature is not an option for the now and future generations, the optimal balance between anthropization, ecological welfare, and biodiversity wealth has to be found. We believe that practical and operational solutions based on AI technologies can help decision-making in this context and facilitate the necessary transitions.

In our project, we revisit several fundamental problems in agroecology using data science and machine learning under the lens of  coviability and its multiple dimensions. Several work have already considered reinforcement learning applied to agroecology: e.g., for irrigation scheduling (Yang et al., 2020),  for optimizing fertilizer usage as well as enhancing the yield (e.g., CropGym environment (Overweg et al., 2021)), or for watershed management (Mason et al., 2016) and biological environment (Binas et al., 2019). However, to the best of our knowledge, none has been proposed to develop actionable research products for monitoring coviability and discovering transition pathways towards coviability. In what follows, we enumerate our project's objectives in addressing four scientific challenges and shortly present our case studies with the application of ML techniques.

\noindent{\bf Challenge 1. Modeling coviability indicators from multimodal, multisource, multiscale data fusion.} As the volume and variety of socio-environmental data and digital resources have dramatically grown in the last decade, it has become crucial to: (1) Align, interconnect, integrate all relevant silos of data at different  spatial and temporal scales local, regional, territorial) as illustrated in Fig. 1: from satellite images, pictures from visitors, data collected through large-scale citizen science platforms such as Pl@ntNet\footnote{\tiny\url{https://plantnet.org}} (Go\"{e}au et~al., 2013), eBird\footnote{\tiny\url{https://ebird.org}}, iNaturalist\footnote{\tiny\url{https://www.inaturalist.org}}   or MoveBank\footnote{\tiny\url{https://www.movebank.org}}; (2) Analyze and explore interrelationships across a variety of disciplines; (3) Develop relevant indicators for assessing vulnerability in socio-ecological systems; and (4) Present the indicators via dashboards and adequate visualization artifacts to the actors, e.g., ad-hoc doughnut representations (Raworth, 2017). In this context, the integration of heterogeneous and multimodal data, preparation, and featurization are challenging tasks.

\noindent{\bf Challenge 2. On-site identification of plants, animals, and complex scenes for computing coviability indicators.} The impacts of climate changes and human activities on biodiversity, ecosystem functioning and environmental resources are numerous, complex, distributed differently according to temporal and spatial scales. The complex and likely reciprocal causal relationships between humans and the environment must be understood according to systemic, multi-scale, and transdisciplinary approaches integrating and interconnecting a large variety of datasets from ecology, remote sensing, local authorities, etc. Given the data uncertainties and what we know about the ecosystem of a site, the analysis of biodiversity remain challenging for automatically identifying multiple plants and animal species as well as decomposing the landscape elements precisely. State-of-the-art techniques allowing to identify plants and animals from individual pictures (e.g., Pl@ntNet or iNaturalist applications) cannot be straightforwardly adapted to our framework and need to be deeply revisited to capture the complexity of our domain to compute local coviability indicators with uncertainty quantification at every data processing step.

\noindent{\bf Challenge 3.  Finding coviability pathways from the diagnostic to a target state.} Understanding how to transition from the current state of coviability to a state that reconciles (i) biodiversity conservation and ecosystem functioning, (ii) social welfare, and (iii) economic resilience is another major scientific issue of our project. As shown in Fig. 1, depending on the profile of the actors (e.g., farmers, indigenous populations, administrators) as the main users of our technology, the considered scale, the constraints associated to the  territory (e.g., economic viability, presence of species close to extinction), and a coviability target co-specified with the actors, an optimal trajectory need to be discovered to support adequate governance schemes. Due to the inherent critical nature of adopting coviability pathways, it is not foreseeable to blindly provide a ranking of trajectories without taking into account the context and multiple alternative spatial, temporal, and actor-dependent viewpoints. To keep the experts and local actors in the loop, we take into account their feedback at every step: from the flora/fauna scene identification/validation, the choice and composition of coviability indicators, the design of dashboards and results of  recommendation, scoring, and explanation of discovered optimal pathways. We revisit this challenge as an optimization problem and use causal reinforcement learning and various optimization methods. Choosing the deep RL architecture as well as the design of relevant coviability indicators are parts of our investigation. Another challenging problem  relates to the study of constrained trajectories (or pathways) to take into account the context.

\noindent{\bf Challenge 4. Transfer to the society and educational programs.} Designing coviability indicators is not politically and socially neutral. Therefore, it is important to involve civil society in the process of data collection, annotation, and technology validation via participatory science: farmers, indigenous populations, and children of school age will be involved in the analysis and interpretation of the data and results (in terms of coviability indicators and transition pathways) since these can often be considered as politico-ecological. We have started to define educational programs in collaboration with schools and universities for transferring the results of our 5-year project and started the development of didactic tools intended for children, teenagers, students (future citizens), teachers, and managers of natural spaces. Pilot demonstrators will help in the ML technology transfer to the society with the challenge to build trust and engage the population.

\section{ML Applications to Four Case Studies }

Figure 1 illustrates our transversal research methodology which we intend to apply and validate on four case studies summarized hereafter. 
{\bf 1) Para\'iba Case Study.}  
At the Northern edge of the Atlantic Forest biome, our project will consolidate the proposal of the Northern Atlantic Forest Agroecological Corridor (NAFAC), bridging science and governance to transform an environmentally degraded and socially vulnerable landscape into a coviable multiscape (i.e., a landscape planned to be multifunctional), providing optimal solutions for conflicting demands between eco-evolutionary and socioeconomic process at the regional scale. In doing so, we expect to enhance the biological connectivity of biodiversity hotspots and promote agroecological transitions and social welfare in rural communities. \\
{\bf Our Goals:} (1) Co-construct, with local communities and managers, sustainability diagnoses depicting the current state of key environmental, social and economic indicators, mapped to large socio-ecological datasets for classify the diagnosis into a coviability scale; (2) Model alternative agroecological transition scenarios using deep RL at the local scale, based on the goals defined by each community involved in the NAFAC proposal; (3) Develop interactive maps, dashboards, explainable classification, and social network tools at the regional scale to support participatory decision-making on spatial strategies fostering ecological and economic connectivity, biodiversity restoration and sustainable production in the NAFAC region.

{\bf 2) Cear\'a Case Study.}  The state of Cear\'a  is the Brazilian state with the largest ratio of area in the Polygon of Droughts. It historically faces the climatic and social effects of drought as the semi-arid climate covers more than 90\% of its territory. Because of the high intra-annual and inter-annual variability of precipitation, the high evaporation rates, the intermittent fluvial regime and the limited underground resources, Cear\'a  often faces water scarcity. The information system and water resources in Cear\'a are managed at the state level. This context reveals the importance of strengthening territorial water governance. 
The integrated knowledge of territorial dynamics at the local level could enable the construction of feasible pathways with multi-objective optimization to optimize water distribution, reduce energy and carbon footprint, and ensure sustainable development.\\
{\bf Our Goals:} (1) Carry out a diagnosis of territorial water management and agriculture by extracting and merging available data to build a knowledge base; (2) Provide the territorial management systems with tools to integrate and process local data and support regression-based predictions for greater efficiency of local water governance and agroecological practices; (3) Provide counter-factual analyses of the prediction results with local what-if scenarios.

{\bf 3)  Rio Grande do Norte Case Study.} 
The Caatinga is the only exclusively Brazilian biome, which is highly threatened by anthropogenic activities. The increasing deforestation rates across the Caatinga are triggering desertification. Developing strategies to boost the restoration of Caatinga's biodiversity and related ecosystem functions, while promoting the socio-productive inclusion of human populations facing social vulnerability, is the key to promoting socio-ecological coviability. This case study benefits from data, expertise, and insights from vegetation restoration experiments performed by UFRN Laboratory of Restoration Ecology at the A\c cu National Forest since 2016, which have received wide international recognition\footnote{\tiny\url{https://ufrn.br/en/press/headlines/38360/esperanca-na-caatinga}}. Our project will co-fund the Centre of Reference for the Caatinga Restoration (CIRCA) in the A\c cu region. It intends to become a reference for biodiversity-based value chains that are dependent on restoration models and supported by local cooperative initiatives.  CIRCA structure is already partially implemented and we expect to implement a pilot-project involving women facing social vulnerability in the development of biodiversity-based products. Although the products to be produced will be decided in partnership with the local women, the restoration plots have shown potential to produce honey, cosmetic inputs, and medicinal plants. We will contribute to the planned observatory of coviability transitions by providing niche models of potential distributions of 600 native plant species that are suitable for use in ecological restoration associated to socio-productive inclusion projects under future climate scenarios. \\
{\bf Our Goals:} (1) Identify, in partnership with local women, socioeconomically viable biodiversity-based products;  (2) Use ML,  existing niche models, population projections, and domestic and international market trade information to predict the short-term and future viability of biodiversity-based value chains; (3) Provide predictions for the geographic variation of ideal plant assemblages ecologically suitable for the restoration of Caatinga biome under current and future climate conditions that will optimize locally the balance between functional diversity of plants and their bioeconomic potential.

{\bf 4)  Pernambuco Case Study.} 
In several regions across the Northeast of Brazil, the ancient indigenous tradition of eating the queens (“i\c c\'as”) of leaf-cutting ants (Atta spp.) remained a relevant dietary item for human populations. Leaf-cutting ants are widely known as critical agricultural pests, imposing severe
economic losses to farmers. Because of
the growing use of pesticides in Brazil, the persistence of the traditional “i\c c\'as” consumption is highly threatened. Nevertheless, the extent to which the consumption of Atta queens by humans can act as a biological control mechanism, mitigating their negative effects on agriculture, remains largely unknown. This case study will investigate the use of leaf-cutting ants as food sources by human populations and how a coviability approach can simultaneously mitigate the ants’ effects on agriculture while maintaining the traditional consumption of these insects.\\
{\bf Our Goals:} (1) Understand the socio-cultural processes that have maintained the traditional consumption of Atta ants alive in the Catimbau region, while this is a declining practice in Brazil; (2) Use deep RL and simulation to plan vegetation restoration plots surrounding agricultural fields to provide alternative resources for Atta ants and sustain alternative biodiversity-based ecosystem services and economical alternatives to local farmers; (3) Map the geographic distribution of edible insects and other species used as food sources, contributing to the identification of potential links between ecosystem restoration and traditional knowledge and practices.

\section{Conclusion}
Understanding and predicting fundamental relationships between socio-economic processes shaping ecological vulnerability and its implications for ecosystem functioning at various spatiotemporal scales is a key universal challenge we shall face in the years to come in order to develop viability strategies that effectively reconcile nature, societies, and economies. A first step is to unravel fundamental socio-ecological processes shaping tensions and antagonisms among social, economic and ecological entities at different temporal and spatial scales. Our project aims at helping draw up coviability diagnoses of a territory by integrating and exploiting various multimodal datasets, predicting  coviability scores using ML, and finding socio-ecological transition pathways using RL. Our research is applied to four case studies across the North-East region of Brazil and is intended to provide actionable ML-based solutions for coviability monitoring and prediction. It combines humanities and social sciences, agroecology, and data and ML science.

\section{Acknowledgement}
This work was supported by RLGR -- Funda\c c\~ao de Apoio \`a Pesquisa do Estado da Para\'iba -- FAPESQ/PB, Grant Number: 3087/2021.

\label{sec:reference}
\nocite{*}
\small
\bibliography{aaai22}
\end{document}